\documentclass[showpacs,reprint,aps,prl,twocolumn,groupedaddress,showpacs,superscriptaddress]{revtex4-1}
\usepackage{graphicx}
\usepackage{dcolumn}
\usepackage{mathrsfs}
\usepackage{epsfig}
\usepackage{subfigure}
\usepackage{color}
\usepackage{sidecap}
\usepackage{float}
\usepackage{bm}

\newcommand{\calL}{{\cal L}}

\begin{document}

\title{Maximum likelihood reconstruction for Ising models with asynchronous updates}

\author{Hong-Li Zeng}
\altaffiliation{Email address: hong.zeng@aalto.fi}
\author{Mikko Alava}
\affiliation{Department of  Applied Physics, Aalto University, FIN-00076 Aalto, Finland}
\author{Erik Aurell}
\affiliation{Department of Computational Biology, KTH-Royal
Institute of Technology, SE-100 44 Stockholm, Sweden }
\affiliation{ACCESS Linnaeus Centre, KTH-Royal Institute of
Technology, SE-100 44 Stockholm, Sweden } \affiliation{Department of
Information and Computer Science, Aalto University, FIN-00076 Aalto,
Finland}

\author{John Hertz}
\affiliation{Nordita, KTH-Royal Institute of Technology and Stockholm University, 10691 Stockholm, Sweden}
\affiliation{The Niels Bohr Institute, 2100 Copenhagen, Denmark}

\author{Yasser Roudi}
\affiliation{Nordita, KTH-Royal Institute of Technology and Stockholm University, 10691 Stockholm, Sweden}
\affiliation{Kavli Institute for Systems Neuroscience, NTNU, 7030 Trondheim, Norway}

\begin{abstract}
We describe how the couplings in an asynchronous kinetic Ising model can be inferred. We consider two cases, one in which we know both the spin history and the update times and one in which we only know the spin history. For the first case, we show that one can average over all possible choices of update times to obtain a learning
rule that depends only on spin correlations and can also be derived from the equations of motion for the correlations. For the second case, the same rule can be derived within a further decoupling approximation. We study all methods numerically for fully asymmetric Sherrington-Kirkpatrick models, varying the data length, system size, temperature, and external field. Good convergence is observed in accordance with the theoretical expectations.
\end{abstract}
\pacs{05.10.-a,02.50.Tt,75.10.Nr}
\maketitle

{\em Introduction.}--- Inferring interactions between the elements of a network can be posed as an inverse
problem in statistical physics either in terms of equilibrium models  \cite{Kappen98,*Tanaka98,*RoudiTyrchaHertz09PRE,*RoudiAurellHertz09,*Aurell12,Bialek06,*Shlens06,*Cocco09,Weigt09}, or non-equilibrium ones.
The latter has recently gained a lot of attention because of the wider generality and relevance to systems where one has data on the system over time
\cite{Pillow08,*RoudiHertz11,*RoudiHertz11JSAT,*Mezard11,*mastromatteo2011,*Tyrchaetal2013,*Zhang2012,Zeng11}.

In this connection, the asynchronous kinetic Ising model offers a powerful platform for
theoretical insight and practical applications. Under detailed balance (symmetric couplings), it converges to the celebrated maximum entropy equilibrium Ising distribution \cite{Glauber63}, that is, the asynchronous model {\em includes} as a subclass the Gibbs equilibrium Ising model.  In many recent works, this equilibrium model is used for inferring functional connectivity and building statistical descriptions, e.g. for neuronal spike trains \cite{Bialek06,*Shlens06,*Cocco09}. However, spike trains and many other real life data come in the form of time series. Since it is only under strict detailed balance that the asynchronous Ising model converges to the equilibrium Ising distribution, it is important to find the relation between the couplings found from the asynchronous model and those from the equilibrium Gibbs distribution. This becomes particularly important for analyzing data using fine time bins at which temporal correlations become important.

The asynchronous Ising model is also important from another perspective. Most of the work on the subject so far
has focused on models with only one {\em type} of stochastic variables. The asynchronous Ising model, however, can be viewed as a doubly stochastic model where in addition to spin configurations, the update times of
the spins are themselves stochastic variables. This differs from the synchronously updated
model where all spins are updated at all times, making the spin configurations the only
stochastic variables \cite{RoudiHertz11}. Doubly stochastic processes are in fact abundant in real life.
An example is a securities market \cite{Ranaldo04,*Maslov2000}  where traders place limit orders: conditional
offers to buy securities if their market price falls below a threshold, or to sell if the market price rises above it.
If offers are made, other traders may respond or not; if they do, transactions take place. Whether or not limit offers are placed define a first set of stochastic variables depending on which transactions may or may not occur, defining a second set.

The presence of two stochastic degrees of freedom raises a number of
questions. How can we infer interactions if the data only contain the history of one of them e.g. the transaction times?
How does this compare with the case where everything is known? When do the two scenarios converge?
Here, starting from two likelihood functions for the data, one
in which update times are known, the other not, we derive two different
learning rules. We show that these learning rules have different precisions for inferring the couplings,
and that they have a nontrivial relation to each other: averaging over possible update
times, they both lead to a third one, but with
different learning rates. Surprisingly, this third learning rule can be also derived
from the forward equations of motion for the correlations of the
asynchronous Ising model \cite{Glauber63} and without appealing to
a likelihood function. This relates two previously unrelated approaches of learning the
couplings. Applying the averaged rule to data from retinal ganglion cells, we find that the connections of the
effective asynchronous model are nearly identical to those of the equilibrium Ising model.
Since the learning rules we derive, as opposed to those for the equilibrium Ising model, do not require calculating a partition function and Monte Carlo sampling, the asynchronous
model offers a much faster way of inferring functional connectivity.

{\em Kinetic Ising model with asynchronous updates.}---  Consider $N$ binary spins, $s_i=\pm 1$, $i=1\cdots N$,
coupled to each other through a matrix $J_{ij}$ and each subject to an external field $\theta_i$. The coupling
matrix need not be symmetric and, consequently, the system may not possess a Gibbs equilibrium state
\cite{Gillespie77}. One can describe this stochastic dynamical system in either of two
ways:

({\bf 1}) Consider a time discretization with steps of size $\delta t$. At each step,
update spin $i$ with probability $\gamma_i \delta t$, where $\gamma_i$ are constants with
dimension of inverse time. We assume $\gamma_i$ to be known {\em a priori}, not a
parameter of the model to be determined. For simplicity, we also assume $\gamma_i=\gamma$ for all $i$
but all our derivations follow in the general case as well.  By ``update'' we mean assigning a new value
$s_i(t+\delta t)$ with probability $\left(1+s_i(t+\delta t)\tanh H_i(t)\right)/2 =
\exp(s_i(t+\delta t)H_i(t))/2\cosh H_i(t)$, where $H_i(t)=\theta_i+\sum_j J_{ji} s_j(t)$
is the total field acting on spin $i$ at time $t$. Of course, the
new value, $s_i(t+\delta t)$ may be equal to the old one; updating does not necessarily
mean flipping.  Multiple spins can be updated in one time step, but for $\delta t \ll 1$ (the
limit we consider) in most steps at most one spin is updated. The synchronously-updated
model is recovered when $\gamma \delta t = 1$.  Thus, one can interpolate between the
synchronous and asynchronous models by varying $\gamma$.  In this formulation, the model
is doubly stochastic: the dynamics of one set of stochastic variables (the spins) are
conditional on the dynamics of the other (the updates). Here we set the
temperature that conventionally appears in this model equal to $1$,
because it can be absorbed into the definitions of the fields and couplings.
Equivalently, our fields and couplings are in units of temperature.

({\bf 2}) Start from the Glauber master equation \cite{Glauber63}.  Then at every step every
spin is flipped with a probability $\gamma \delta t \left(1-s_i(t)\tanh H_i(t)\right)/2$.
As in scheme (1), multiple spins can flip in a single time step, but this happens
with probability of order $(\delta t)^2$. Thus, $\delta t \ll 1$, in most time intervals at most one spin is flipped.

The difference between the schemes is that in scheme (1) we  have two sets of random variables, the
update times (which we denote by $\{\tau_i\}$) and the spin histories $\{s_i(t)\}$, while
scheme (2) contains only the $\{ s_i(t) \}$. One can easily show that marginalizing
out the $\{\tau_i\}$ in scheme (1) leads exactly to scheme (2),
even if $\gamma \delta t$ is not small.  Thus, all
averages over histories involving spins only (i.e., not involving the update times) will
be the same in the two schemes. Nevertheless, knowing ``the history of the
system'' (i.e., a realization of its stochastic evolution) means something different in
the two schemes.  In the first we know all the update times, while in the second we only
know those at which the updated spins flipped.   We will see below that knowing these
extra data influences the performance in reconstructing the couplings.   Which scheme is relevant for inferring the couplings from data depends on the
specific nature of the system being modeled and the data available.  The ``update times"
may be meaningful and, if so, available in some cases and not in others.

{\em Two likelihoods to maximize.}---
Consider scheme (1) above.  Suppose we are given a history of the system,
i.e., the data $s \equiv \{ s_i(t) \}$ and $\tau \equiv \{ \tau_i \}$,  of length $L =
T/\delta t$ steps, and we are asked to reconstruct the couplings and fields.  We
do this by maximizing the likelihood $P(s,\tau) = P(s | \tau)p(\tau)$ over these
parameters.  For each spin $i$, the $\tau_i$ are a (discretized) Poisson process, i.e.,
every $t$ has probability $\gamma \delta t$ of being a member of the set $\tau$.  Thus the
probability of the update history, $p(\tau)$, is independent of the model parameters, and
we can take as objective function $\log P(s | \tau)$, i.e.,
\[ \calL_1  = \sum_i \sum_{\tau_i}\left[ s_i(\tau_i+\delta t)H_i(\tau_i) - \log 2 \cosh H_i(\tau_i)\right].\]
This is just like the synchronous-update case except that the sum over times is only over
the update times.  It leads to a learning rule
\begin{equation}
\delta J_{ij} \propto \frac{\partial
\calL_1}{\partial J_{ij}}=
\sum_{\tau_i} [s_i(\tau_i+\delta t)-\tanh(H_i(\tau_i))]s_j(\tau_i).   \label{eq:der2}
\end{equation}
Defining $J_{i0} = \theta_i$, $s_0(t) = 1$, this equation also includes the learning rule for $\theta_i$.  We call this algorithm ``spin- and update-history-based'', or
``SUH''.

In scheme (2), we know only the spin history, not the update times.
Since this scheme is equivalent to the first one with the $\tau_i$ marginalized out,  we
treat it by maximizing $P(s) = \sum_{\tau} P(S|\tau)p(\tau)$ \cite{kipnis99}, leading
to
\[\calL_2=\sum_{i,t}\log \left[(1-\gamma \delta t)\delta_{s_i(t+\delta t),s_i(t)}+\gamma \delta t \frac{{\rm e}^{s_i(t+\delta t)H_i(t)}}{2\cosh H_i(t)}\right].\]
as objective function. Separating terms with and without spin flips,  the resulting learning rules will be
\begin{eqnarray}
\delta J_{ij} &\propto& \frac{\partial \calL_2}{\partial J_{ij}}= \sum_{\rm flips} [s_i(t+\delta t)-\tanh(H_i(t))]s_j(t)          \nonumber \\
&+& \frac{\gamma \delta t}{2}\sum_{\rm  no \hspace{2 pt} flips} q_i(t) s_i(t+\delta t)s_j(t),
\label{eq:der}
\end{eqnarray}
where $q_i(t)\equiv [1-\tanh^{2}(H_i(t))]$, and it includes the rule for the $\theta_i$ with the convention $J_{i0} = \theta_i$, $s_0(t) = 1$.
We call this the ``spin-history-only'' (``SHO'') algorithm.

Reconstruction errors for both algorithms can be calculated by analyzing the Fisher information matrices.
For SHO the Fisher matrix elements read
\begin{eqnarray}
&&-\frac{\partial^2\calL_2}{\partial J_{ij} \partial J_{kl}}=\delta_{ik} \sum_{\rm flips} q_i(t) s_{j}(t) s_l(t)    \label{eq:fL1}  \\
&&+2\delta_{ik}\gamma \delta t \sum_{\rm  no \hspace{2 pt} flips} q_i(t)  s_i(t+\delta t)\tanh(H_i(t)) s_{j}(t) s_l(t) .   \nonumber
\end{eqnarray}
In the weak coupling limit, this matrix has nonzero elements only for $j=l$,
and the mean value of these non-zero elements yields the inverse of
the mean square reconstruction error (MSE). Without external fields, the
second term in Eq. (\ref{eq:fL1}) vanishes; thus, the MSE
in this case is $2/(T\gamma)$, noting that the probability
that a time step is a flip is $\gamma \delta t/2$. For SUH the calculation is analogous and
for $\theta_i=0$ and weak couplings, the MSE will be $(T\gamma)^{-1}$, i.e., a factor of two smaller than for SHO.

{\em History-averaged learning.}--- SUH and SHO utilize explicitly their respective full
model histories, both $\{s_i(t)\}$ and $\tau_i$ for SUH and $\{s_i(t)\}$ for SHO.
Below we derive a third rule by averaging the one for SUH, Eq.\ (\ref{eq:der2}), over all
update histories.  Defining $C_{ij}(t) \equiv \langle s_i(t_0+t)s_j(t_0)\rangle$, we have
\[ {\dot C}_{ij}(t)=\lim_{\delta t \to 0}\frac{\langle s_i(t+\delta t)s_j(t_0)\rangle -\langle s_i(t)s_j(t_0)\rangle}{\delta t},\]
where $\langle \cdots \rangle$ means an average over all realizations of the stochastic dynamics.
Separating time steps into those at which an update occurred and those at which no update occurred yields
\[ {\dot C}_{ij}(t)= \lim_{\delta t\to0} \left\{\gamma\delta t\frac{\langle s_i(\tau_i+\delta t)s_j(t_0)\rangle_{\tau_i}-\langle s_i(\tau_i)s_j(t_0)\rangle_{\tau_i}}{\delta t}\right\}\]
There is no contribution from steps with no flip because then $s_i(t+\delta t) = s_i(t)$ and the numerator would be zero.  Thus we have expressed the average
over all realizations of the first term in Eq.\ (\ref{eq:der2}) in terms of spin correlation functions and their time derivatives:
\begin{equation}
\langle s_i(\tau_i +\delta t)s_j(\tau_i) \rangle_{\tau_i} = \frac{1}{\gamma} {\dot C}_{ij}(0)+C_{ij} (0).
\label{cpluscdoteqn}
\end{equation}
In averaging the second term in Eq. (\ref{eq:der2}), the average over $\{\tau_i\}$ can be replaced by an average over all times, since the quantity $\tanh H_i(t)s_j(t)$ is insensitive to whether an update is being made.  Thus, averaging Eq. (\ref{eq:der2}) over all possible histories yields
\begin{equation}
\delta J_{ij}\propto \gamma^{-1}{\dot C}_{ij}(0)+C_{ij}(0) -\langle \tanh(H_i(t))s_j(t)\rangle.
\label{eq:J2av}
\end{equation}
We will refer to the update rule given by Eq.\ (\ref{eq:J2av}) as the averaged-SUH rule, or ``AVE'' . This rule has the same structure as the
one for the synchronous-update model \cite{RoudiHertz11}, with $\langle
s_i(t+1)s_j(t)\rangle$ replaced by $C(0) + \gamma^{-1}{\dot C}(0)$.

AVE requires knowing the equal-time correlations, their derivatives at $t=0$, and $\langle \tanh(H_i(t))s_j(t)\rangle$.  This latter quantity depends on the model parameters (through $H_i(t)$), so, in practice, estimating it at each learning step requires knowing the entire spin history, the same data as SHO learning needs.

Can we derive an algorithm like Eq.\ (\ref{eq:J2av}) from SHO learning by averaging over spin flip times in the same way we did by averaging SUH learning over update times?   Denote the local fields at time $t$ generated by the true model (the one that generated the data) by $\tilde{H}_i(t)$, and, as before, the local field calculated using the inferred parameters as $H_i(t)$.  At each time step $t$, then, the probability of flipping spin $i$ is $\gamma \delta t[1- s(t)\tanh \tilde{H}_i(t)]/2$.  We thus have to allot the first term in Eq.\ (\ref{eq:der}) a weight $\gamma \delta t[1- s(t)\tanh \tilde{H}_i(t)]/2$ and the second a weight $1 -\gamma \delta t[1- s(t)\tanh \tilde{H}_i(t)]/2 \approx 1$ getting
\begin{eqnarray}
\delta J_{ij} \propto \left\langle \frac{\partial \calL_1}{\partial J_{ij}}\right\rangle_{0}
&=& \frac{\gamma}{2T}\int dt [\tanh \tilde H_i(t)- \tanh H_i(t)]\nonumber\\
&\times& [1+s_i(t)\tanh H_i(t)]s_j(t).          \label{eq:aalearning}
\end{eqnarray}
The learning thus converges when the discrepancy $\tanh(H(t))-\tanh(\tilde H(t))$ is
zero.  Noting also that the arguments above leading to Eq.~(\ref{cpluscdoteqn}) yields $
\langle \tanh \tilde H(t)s_j(t)\rangle_t =\gamma^{-1} \dot{C}(0)+C(0)$,
we write Eq.~(\ref{eq:aalearning}) as
\begin{eqnarray}
\delta J_{ij}
\propto \gamma^{-1}\dot{C}_{ij}(0)+C_{ij}(0) - \langle \tanh H_i(t)s_j(t)\rangle_t\nonumber\\
+  \langle  [\tanh \tilde H_i(t)-\tanh H_i(t)]  s_i(t)\tanh H_i(t)s_j(t)\rangle_t
\label{AA2}
\end{eqnarray}
The first line is identical to Eq.\ (\ref{eq:J2av}). We can obtain a learning rule heuristically by an {\em ad hoc} factorization of the average in the second line as
$\langle  [\tanh \tilde H_i(t)-\tanh H_i(t)]  s_i(t)\tanh H_i(t)s_j(t)\rangle_t \approx \langle  \tanh \tilde H_i(t)-\tanh H_i(t) s_j(t) \rangle_t \langle s_i(t)\tanh H_i(t)\rangle_t$,
yielding
\begin{eqnarray}
\delta J_{ij} &\propto&  [\gamma^{-1}\dot{C}_{ij}(0)+C_{ij}(0) - \langle \tanh H_i(t)s_j(t)\rangle_t ] \nonumber \\
&\times& \langle [1+s_i(t)\tanh H_i(t)] \rangle_t.
\end{eqnarray}
This just amounts to varying the learning rate in Eq.\ (\ref{eq:J2av})
proportional to the time-averaged probability of not flipping
according to the model. Thus we arrive by a different route at the AVE rule,
Eq.\ (\ref{eq:J2av}).

We compared the performance of the algorithms
SUH, SHO, and AVE to each other and to
the naive mean-field (nMF) and Thouless-Anderson-Palmer (TAP)
approximations to AVE investigated in \cite{Zeng11} for
fully\- asymmetric Sherrington-Kirkpatrick models
\cite{Sherrington75}. The couplings are zero-mean i.i.d. normal
variables with variance $g^2/N$ ($J_{ij}$
is independent of $J_{ji}$). We study these for different values of $g$ and $\theta$,
the system size $N$ and the data length $L$.  As a performance measure, we use the MSE on the $J_{ij}$.

\begin{figure}[b]
\center
\includegraphics[height=4cm,width=4cm]{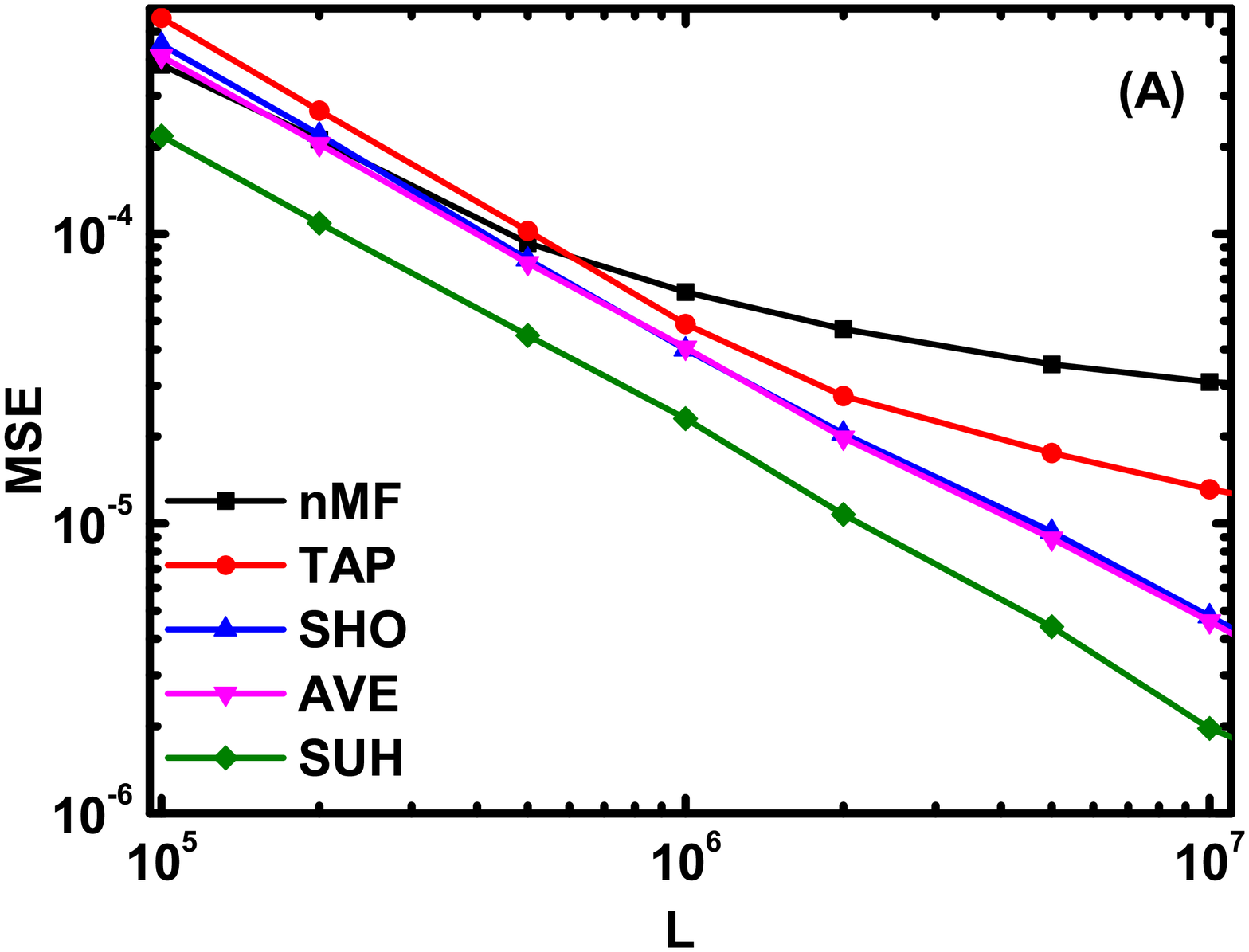}
\includegraphics[height=4cm,width=4cm]{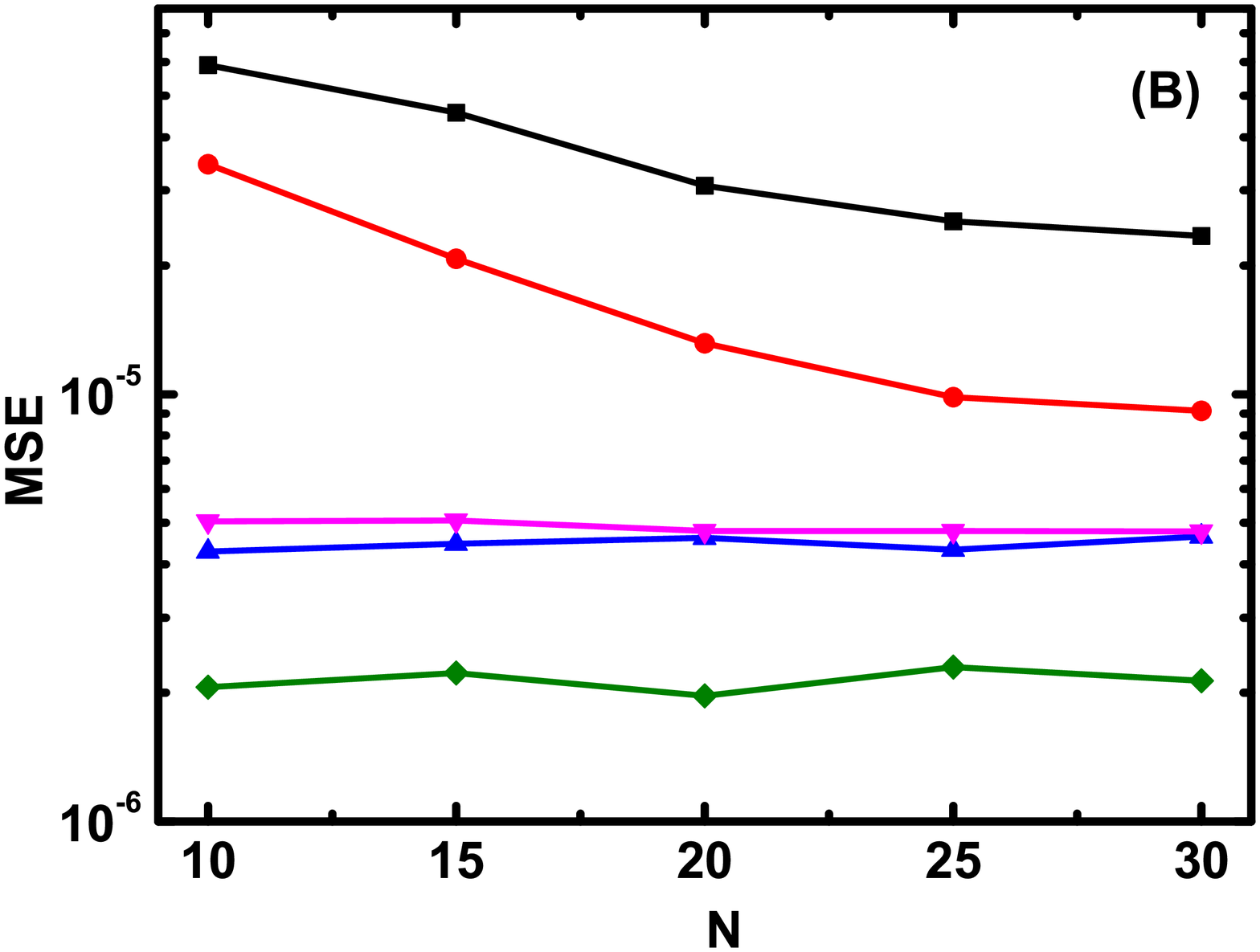}  \\
\includegraphics[height=3.8cm,width=4cm]{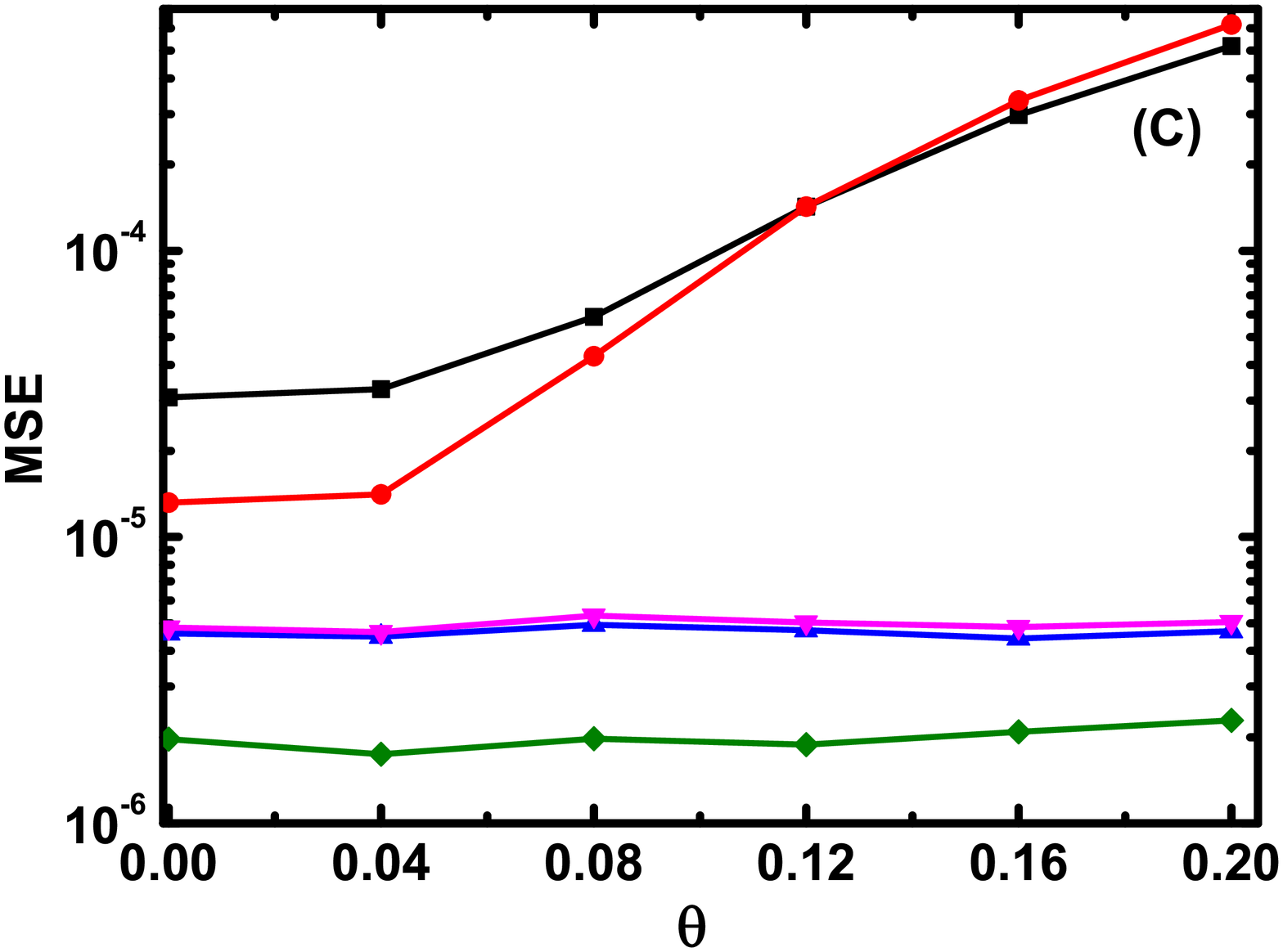}
\includegraphics[height=4cm,width=4cm]{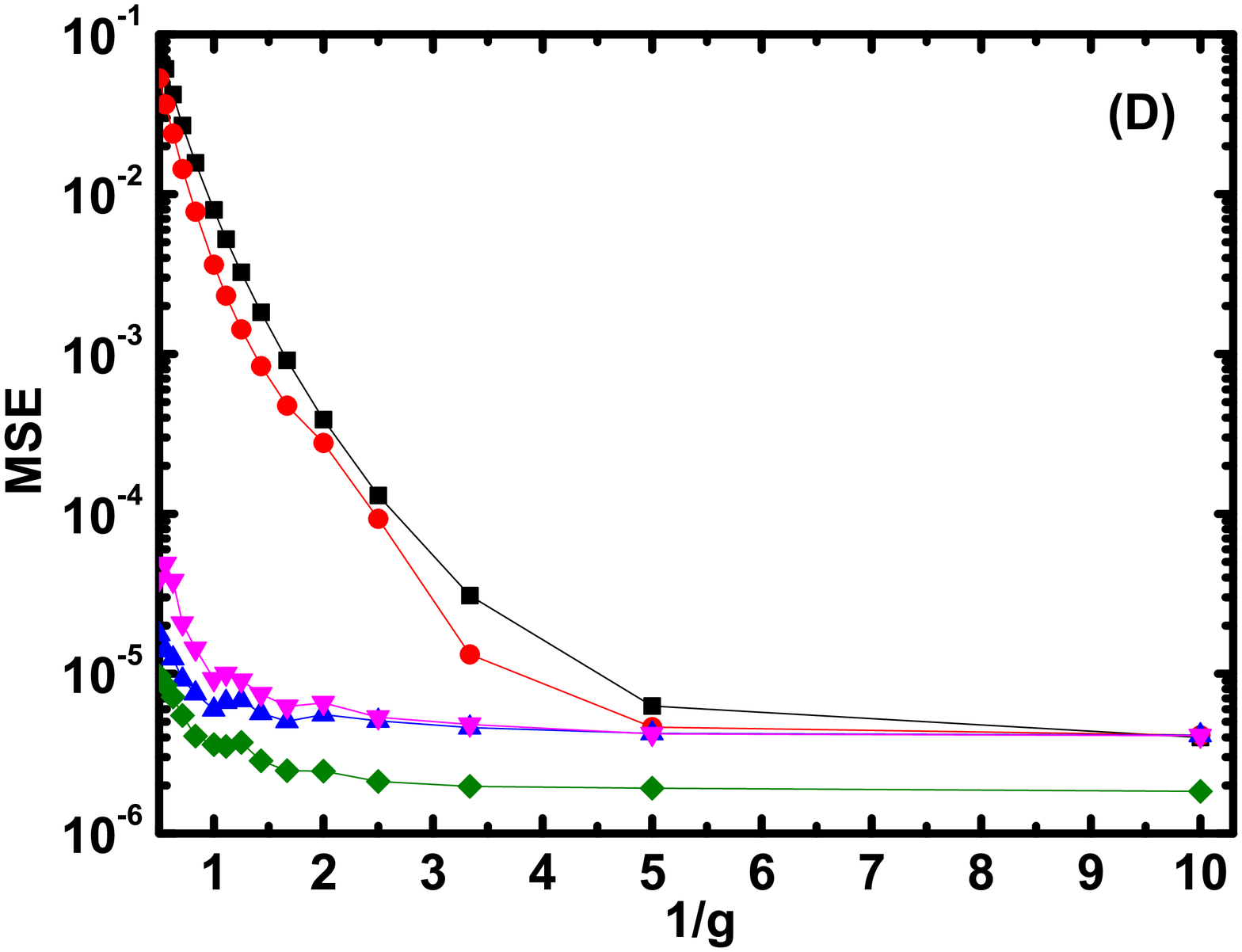}
\caption{(Color online) Mean square error (MSE) versus (A) data length $L$, (B) system size
$N$, (C) external field $\theta$ and (D) temperature $1/g$. Black squares show nMF, red circles, TAP, blue up
triangle SHO, pink down triangle AVE and green diamond SUH, respectively. The
parameters are $g=0.3$, $N=20$, $\theta=0$, L=$10^7$ except when varied in
a panel.}
\label{Fig1}
\end{figure}

Fig.~\ref{Fig1} shows the performance of the algorithms.   As
anticipated above, the error for SUH is half of that for SHO
learning; see Fig.~\ref{Fig1}A.  The same panel
also shows that AVE and SHO appear to perform equally well for large enough $L$.
In retrospect, this is not surprising, since both
algorithms effectively use the same data (the spin history). For small $L$, the averaging that yields AVE from SHO may be prone to
fluctuations yielding the two learning rules behaving differently.
Fig.~\ref{Fig1}B shows that the MSE for the exact algorithms is
insensitive to $N$, while the approximate algorithms improve as $N$
becomes larger (note however the opposite trend in
Fig.~\ref{Fig1}A); in these calculations, the average numbers of
updates and flips per spin were kept constant, taking $L =
5\times10^5N$.)   Fig.~\ref{Fig1}C shows that the performance of
the three exact algorithms is also not sensitive at all to $\theta$, while nMF and TAP work noticeably less well
with a non-zero $\theta$. Finally, the effects of (inverse)
$g$  are depicted in Fig.~\ref{Fig1}D. For fixed
$L$, all the algorithms do worse at strong couplings (large $g$). The nMF and TAP do so in a much more clear fashion
at smaller $g$, growing approximately exponentially with $g$ for $g$
greater than $\approx 0.2$.   In the weak-coupling limit, all
algorithms perform roughly similarly, except that SUH enjoys its
factor-2 advantage (conferred by knowledge of the update times), as
already seen in Fig.~\ref{Fig1}A.

\begin{figure}[t]
\center
\includegraphics[height=3.6cm,width=8cm]{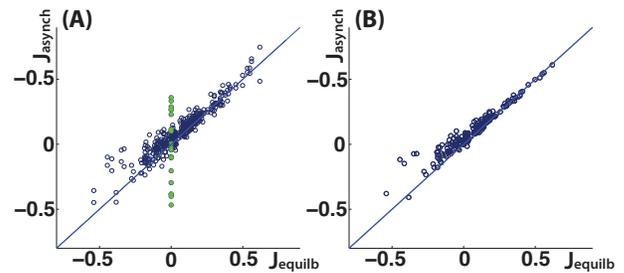}
 \caption{(Color online) Asynchronous versus equilibrium couplings for retinal data. (A) The full asynchronous model. Green squares show the self-couplings which by convention are equal to zero for the equilibrium model. (B) The results when at every iteration the self-couplings were put to zero and the matrix was symmetrized. }
\label{Fig2}
\end{figure}

We applied the learning rule Eq.\ (\ref{eq:J2av}) to spike trains from $20$ retinal ganglion cells and compared the inferred
couplings with those of the Gibbs equilibrium model (see Supp Mat for details). Fig.\ \ref{Fig2}A shows that the Gibbs equilibrium and
kinetic Ising couplings are very similar. Furthermore, the asynchronous model allows the inference of self-couplings (diagonal elements of the coupling matrix)
which are not present in the equilibrium model.  This result provides a rationale for the use of the maximum entropy
equilibrium Ising model: if the asynchronous couplings were very different form the equilibrium ones, it would have
meant that the real dynamical process did not satisfy the Gibbs equilibrium conditions and that the final distribution of
states is not the Gibbs equilibrium Ising model. In fact, we also tested what happens to the couplings of the asynchronous
model if during learning we symmetrized the couplings matrix at each iteration by adding its transpose to itself and dividing
by two and also putting the self-couplings to zero. Fig.\ \ref{Fig2}B shows that the resulting couplings now get even closer
to the equilibrium ones.  Since inferring the equilibrium model is an exponentially difficult problem, requiring time consuming
Monte Carlo sampling, these results have an important pragmatic consequence for inferring retinal functional connectivity.
This is because the asynchronous approach does not require Monte Carlo sampling: the averages on the right hand side of
Eq.\ \ref{eq:J2av} are all over the data.  The asynchronous learning rules thus allow the inference of functional connections
that for the retinal data largely agree with the maximum entropy equilibrium model, but the inference is much faster.

{\em Discussion.}--- A surprising observation is that Eq.\ (\ref{eq:J2av}) that we derived by maximizing the likelihood, can also be derived from
a totally different route. For a kinetic Ising model, the equation of motion for the
correlations given $\theta$ and $J$ is
$\gamma^{-1}\dot{C}_{ij}(0)+C_{ij}(0)=\langle \tanh H_i(t)s_j(t)\rangle_t$ \cite{Glauber63}.
This equation holds for correct couplings, and thus a heuristic learning is given by just
adjusting the couplings proportional to the difference of the two sides. This again yields
Eq.\ (\ref{eq:J2av}), and the linearized version of it would, in fact, be the mean-field
inference algorithm for the asynchronous model used in \cite{Zeng11}. Our results show
that this rule is not merely heuristic: it can be derived starting from the likelihood of
the data, whether assuming that update times are known or not, and averaging over the
update times.

Here we addressed the problem of inferring the couplings in a non-equilibrium
system: the asynchronous, asymmetrically coupled kinetic Ising model.  We showed how
to derive three different learning algorithms, utilizing three different levels of detail
of the history of the system: the full spin and update history, the spin history only, and
spin correlations at and near $t=0$ only. The methods show performance that is
promising in practical terms, agrees with theoretical expectations, and in particular is
superior to approximate methods found earlier. We expect that the reasoning behind
our results on deriving and relating different learning rules can be extended to a variety
of inverse statistical mechanics problems beyond the particular case of the kinetic Ising
model.

{\em Acknowledgements.}--- This work has been supported by the Finnish graduate school for Computational Science
(FICS), the Academy of Finland as part of its Finland Distinguished Professor
program project 129024/Aurell and the Centers of Excellence COMP and COIN, as well as NORDITA and the Kavli Foundation. The
authors acknowledge Manfred Opper for discussions and Michael Berry for providing the retinal data.
\bibliography{MAExactLearning}
\end{document}